\documentclass[%
reprint,
 unsortedaddress,
 amsmath,amssymb,
 aps,
]{revtex4-2}
\usepackage{caption}
\pdfoutput=1
\usepackage{subcaption}
\usepackage{graphicx}
\usepackage{dcolumn}
\usepackage{bm}
\usepackage{cleveref}
\usepackage{tablefootnote}
\usepackage{xcolor}

\begin{document}

\preprint{APS/123-QED}

\title{Kinetics of hydrogen and vacancy diffusion in iron: A Kinetic Activation Relaxation technique (k-ART) study}

\author{Aynour Khosravi}
 \email{aynour.khosravi@umontreal.ca} 
\affiliation{Département de Physique and Regroupement québécois sur les matériaux de pointe,\\ Université de Montréal, Montréal, Canada.}

\author{Jun Song}
 \email{jun.song2@mcgill.ca}
\affiliation{Department of Mining and Materials Engineering, McGill University, Montréal, Canada.}

\author{Normand Mousseau}
 \email{normand.mousseau@umontreal.ca}
\affiliation{Département de Physique and Regroupement québécois sur les matériaux de pointe,\\ Université de Montréal, Montréal, Canada.}

\date{\today}

\begin{abstract}
We investigate hydrogen (H) and mono and divacancy-hydrogen complexes (VH$_x$ and V$_2$H$_x$) diffusion in body-centered-cubic (BCC) iron using the kinetic Activation-Relaxation Technique (k-ART), an off-lattice kinetic Monte Carlo approach with on-the-fly event catalog building, to explore diffusion barriers and associated mechanisms for these defects. 
K-ART uncovers complex diffusion pathways for the bound complexes, with important barrier variations that depend on the geometrical relations between the position of the inserting Fe atom and that of the bound H. Since H is small and brings little lattice deformation around itself, these bound complexes are compact, and H is fully unbound at the second neighbor site already. As more H are added, however, vacancies deform and affect the lattice over longer distances, contributing to increasing the VH$_x$ complex diffusion barrier and its impact on its local environment. We find, moreover, that the importance of this trapping decreases when going from mono to divacancy complexes, although diffusion barriers for these complexes increase with the number of trapped H. 

\end{abstract}

\maketitle


\section{\label{sec:level1} Introduction}

The degradation of metals' structural properties by H, called H embrittlement (HE)~\cite{nagumo2016fundamentals} is a phenomenon that has long been a concern for a number of industrial sectors~\cite{johnson1875ii}, including the aerospace and fasteners industries. Yet, despite tremendous research efforts, it remains poorly understood to this day. High-strength steels are particularly vulnerable to embrittlement because the high mobility of H facilitates insertion and aggregation~\cite{oriani1978hydrogen}. Not surprisingly, therefore, HE's strong industrial relevance and scientific merits have resulted in significant research endeavours  and the proposal of multiple mechanisms that aim to unravel the fundamental physical processes underlying it~\cite{myers1992hydrogen,wipf1997hydrogen,pundt2006hydrogen}.

In general, it is assumed that HE causes crack initiation followed by progressive crack propagation when diffusible (mobile) H migrates from the bulk lattice to stress concentrations where H accumulates the most. Alloys such as steel and iron-based alloys are predominantly characterized by their chemical composition, which includes the distribution of carbon (C), as well as their microstructure, grain boundary structure, dislocation density, and vacancy clustering and concentration~\cite{nagumo2016fundamentals}. To occur HE, requires a sufficient concentration of mobile H, enough relaxation time to allow H diffusion, and stress concentration at deformations of a certain type. These conditions lead to crack initiation/propagation at applied stress ranges far below the yield strength of the material~\cite{pundt2006hydrogen}.
To explain these requirements and various observations, numerous models of HE have been proposed, such as the hydrogen enhanced localized plasticity (HELP), the hydrogen enhanced decohesion mechanism (HEDE), the hydrogen enhanced strain-induced vacancy formation (HESIV), and the adsorption-induced dislocation emission (AIDE)~\cite{nagumo2016fundamentals}. HELP, for example, proposes a mechanism in which hydrogen promotes dislocation motion, and local plastic deformation results~\cite{martin2019enumeration}. For its part, the HESIV model suggests that hydrogen enhances the formation of clusters of strain-induced vacancies which destabilize the local plastic deformation~\cite{nagumo2019predominant,nagumo2004hydrogen}. Furthermore, research on carbon and hydrogen interaction with vacancies demonstrated that the point defect clusters in Fe-C-H decoupled into binary systems Fe-C and Fe-H effectively~\cite{monasterio2009hydrogen}. Consequently, even at low hydrogen concentrations, significant clusters of hydrogen-vacancies are expected. A combined thermal desorption spectroscopy and internal friction study, for example, suggests that carbon reduces vacancy mobility, allowing clustering and growth that traps hydrogen~\cite{vandewalle2022combined}.

In all cases, a key underlying factor of HE is the interaction between H and microstructural heterogeneities in a material, such as vacancies, dislocations, and grain boundaries. Yet, due to the low mass and high diffusivity of H,  accurate H mapping in microstructures at the atomic scale is very challenging. It is therefore possible to reach severe premature failure at atomic levels of H which are difficult to detect in experiments~\cite{koyama2017recent}.
In Fe-based steels, high-resolution SEM (scanning electron microscopy) studies~\cite{neeraj2012hydrogen} revealed nanovoids in the presence of hydrogen on the conjugate fracture surfaces of quasi-brittle facets. As well, according to TXM (transmission X-ray microscopy) results~\cite{lee2023detection}, voids at crack tips are elongated and smaller than voids in uncharged samples, with quasi-cleavage fractures and sharper crack tips caused by these voids in the presence of hydrogen. In hydrogen-charged samples, the voids in the crack tip region have an elongated shape, suggesting that growth is inhibited in the loading direction while coalescence is favored.  Another study~\cite{chiari2021strain} used variable temperature positron annihilation lifetime spectroscopy to determine whether hydrogen-induced defects are responsible for hydrogen embrittlement. Defects were observed in hydrogen-charged pure iron when it was deformed at different tensile strain rates while it was at room temperature. As a result, hydrogen-stabilized vacancy clusters accumulate locally in high concentrations in H-embrittled iron. On the theoretical size, analytic thermodynamic model developed by Nazarov, Hickel and Neugebauer~\cite{nazarov2010first} from DFT data showed that the presence of H can enhance the vacancy concentration by a factor of seven in fcc Fe, a possible explanation for  superabundant vacancy (SAV) formation. Interestingly, the study also found that, however big the effect of H on vacancies may be, vacancies have almost no effect on the total H concentration, except under high-temperature and extreme H-rich conditions. 

Going further, understanding how voids contribute to crack growth, which results in mechanisms of nanovoid formation, also requires further investigation. Indeed, spatial H mapping and temporal H tracking through experimental approaches are the two greatest challenges on the path toward a better understanding of HE. As computing power increases and computational methods advance, atomistic simulations are becoming new tools for not only studying the interactions between H and metals but also for proving the proposed mechanisms of HE. Here, with the help of the kinetic activation relaxation technique (k-ART)~\cite{el2008kinetic, beland2011kinetic}, an off-lattice kinetic Monte Carlo algorithm with on-the-fly catalog building, we aim to offer a more detailed characterization of the atomistic details surrounding the relationship between H and vacancy motion in these materials and to provide a basis for addressing this multi-scale challenges.

This paper is structured as follows. We begin by introducing the methods used in this article as well as the simulation details (Section~\ref{section:Methodology}).  Section~\ref{section:Results} shows the migration energies of the hydrogen atom and vacancy in bcc Fe and the effect of hydrogen on vacancy diffusivity. In the last section~\ref{section:Discussion}, we discuss our findings and their implications for hydrogen-assisted defects.

\section{\label{sec:level1}Methodology}
\label{section:Methodology}
\subsubsection{The kinetic activation-relaxation technique (k-ART)}

With constant improvements in methods and computing powers, computational methods and atomistic simulations are playing an ever more important role for understanding the microscopic processes associated with atomic diffusion, in particular H, as direct observation at the appropriate time and length scales are challenging with currently available experimental methods. 

In this study, simulations are first carried out using the method of the kinetic activation-relaxation technique (k-ART)~\cite{el2008kinetic, beland2011kinetic}, an off-lattice kinetic Monte Carlo (KMC) method that uses the activation-relaxation technique nouveau (ARTn) method~\cite{barkema1996event, malek2000dynamics, mousseau2012activation} for generating activated events around specific configurations, and NAUTY, a topological analysis package, for the generic classification of events. 

We briefly go over the basic algorithm of the k-ART method  as well as the parameters. Based on a system relaxed to a local minimum, NAUTY, a topological analysis library developed by McKay~\cite{mckay1981practical}, is used to find local topology for each atom; the generated graph for any atom includes all atoms within a 6~Å radius of the central atom, or around 65 atoms, with vertices connecting atoms within 2.7~Å of each other, which  corresponds here to the first neighbor shell. The constructed connectivity graph is then sent to NAUTY, which returns a unique identifier that characterizes its automorphic group, including chemical identity. It is assumed that all atoms with the same topology have the same list of activated mechanisms; this assumption  is validated for every event and corrected through changes in various cut-offs when this is not the case~\cite{beland2011kinetic}. Here, k-ART identifies 16 topologies for a single H interstitial in perfect bcc structure, including topologies centered on the H and Fe atoms. 

For each topology, a preset number of  ARTn searches are launched to identify events associated with these topologies~\cite{barkema1996event, malek2000dynamics, machado2011optimized}, this number is increased when symmetrical events are found and as a function of the recurrence of a given topology in the system~\cite{beland2011kinetic}.

An ARTn event search consists of the following steps: (i) the local environment surrounding a select atom is deformed in an arbitrary direction, allowing the rest of the system to relax partly, until the lowest eigenvalue of the Hessian matrix, obtained using the Lanczos algorithm, becomes negative, which indicates that the system is outside the initial harmonic well; (ii) a series of pushes are made along the negative curvature, with the force in the hyperplane perpendicular to the negative curvature direction minimized after each push until the total force reaches a predefined threshold value, near zero, indicating that a first-order saddle point has been achieved; (iii) the system is moved over the saddle point and relaxed into a new minimum.

In this work, each new topology is subjected to fifty independent ARTn searches. When an event is entered into the database, it is also added, along with the reserve event, to the binary tree of events and histogram. Once the catalog is fully updated and the tree is completed for the current atomistic configuration, generic events are ordered according to their rate
\begin{equation} \label{eq:1}
\Gamma_i = \nu_{0} exp \left( {-\frac{E_b}{k_B T}} \right)
\end{equation}
where $E_b$ represents the activation (barrier) energy for event $i$; this is measured as the difference between the transition state and the initial minimum. $\nu_0$ denotes the attempt frequency (prefactor). 

A harmonic approximation for this quantity (harmonic Transition State Theory (hTST) ~\cite{vineyard1957frequency,gelin2020enthalpy}) is used, unless specified. The hTST approximation determines the prefactor from the ratio between vibrational frequencies at the initial minimum and at the saddle point. Vibrational frequencies change under local deformations because of interatomic interactions, microstructure and deformation characteristics. It is calculated as follows:
\begin{equation} \label{eq:2}
    \nu_0^{htst}= \frac{\prod\limits_{i=1}^{N}\nu_i^m}{\prod\limits_{i=1}^{N-1}\nu_i^s}.
\end{equation}
$\nu_i^s$ and $\nu_i^m$  represent the vibrational frequencies at the saddle point $\nu_i^s$ and $\nu_i^m$ at the minimum. The products on these frequencies are performed over the real values only (the imaginary frequency at the saddle point is left out of this product).  Vibrational frequencies are obtained through diagonalization of the dynamical matrix, 
\begin{equation}\label{eq:3}
    D_{i\alpha j \beta} = \frac{1}{\sqrt{m_i m_j}}\frac{\partial^2 V}{\partial{x_{i,\alpha}}\partial{x_{j,\alpha}}},
\end{equation}
where indices $i$, $j$ run over  all atoms and $\alpha$, $\beta$ over Cartesian coordinates $(x, y, z)$, $V$ is the interaction potential, and $m_i$ is the atomic mass.
In hTST theory, the transition rate is defined by the temperature T as follows: 
\begin{equation} \label{eq:4}
\Gamma_{is} = \nu_0^{htst} exp \left( {-\frac{E_s - E_i}{k_B T}} \right)
\end{equation}
$E_s$ and $E_i$ represent the saddle point and the initial minimum configurational energies.
Events are sorted from this preliminary rate evaluation and all those with a minimum probability of occurrence (here, 1 in 10~000 and more) are fully reconstructed and re-converged to account for any elastic deformation. Once this is done, the total rate is re-evaluated.
As presented in the result section, we find that the hTST prefactor is relatively constant in this system (within a factor of two or so) and, therefore, we also use a constant prefactor set at $2.2 \times 10^{13}$ Hz in cases that are specified.

KMC time steps are computed according to a Poisson distribution, 
\begin{equation}\label{eq:5}
    t = - \frac{\ln \mu}{\sum_i \Gamma_i},
\end{equation}
with $\mu$ being a uniformly random number distributed number between $[0,1]$ and $\Gamma_i$, the rate of each event accessible by the configuration.

The incremental binding energy of the \textit{x}th H atom in a monovacancy is obtained using 
\begin{equation}\label{eq:6}
E_{B}^{inc} = [E(H_{x-1}V) + E(H_T)] - [E(H_{x} V) + E_0],
\end{equation}
where $E(H_{x-1}V)$ represents the energy of the system with one vacancy +$(x-1) H$ and  $E(H_{x}V)$ denotes the energy of the system with one vacancy + $x H$ atom. $E(H_T)$ is the energy of the system with no vacancy and one H positioned into a tetrahedral interstitial site (T-site). $E_0$ is the total energy of a perfect Fe system. The solution energy of the interstitial H is defined as follows~\cite{PhysRevB.82.235125}:
\begin{equation}\label{eq:solution}
E_{sol}^{int} = E(H_T)- E_0 - \frac{1}{2} E_{H_2},
\end{equation}
where $E_{H_2}$ represents the total energy of the H molecule at 0 K in its equilibrium configuration.
As part of the simulation, we examine the behavior of H in both the perfect BCC iron and the presence of vacancies. Vacancies can act as strong traps for H diffusing through the bulk~\cite{takai2008lattice}. For computational efficiency, we define a vacancy as a lattice site with no Fe atom within 0.5 \AA. These values take into account the fact that KMC simulations do not include thermal displacements and their focus is on minima and first-order saddle points.

\subsubsection{Interatomic potential}

A Mendelev's~\cite{mendelev2003development,ackland2004development} EAM parameterization is most commonly used for FeH in the Fe-H models, complemented with Fe-H and H-H parts. However, the H-H interaction is inadequately described in some of these potentials, resulting in nonphysical clustering of interstitial H atoms~\cite{ramasubramaniam2010erratum}. In this work, LAMMPS' implementation of Finnis-Sinclair type embedded-atom-method potentials with parameters adapted to Fe-H both by Song \textit{et al}~\cite{song2013atomic} and Ramasubramanian \textit{et al.}~\cite{ramasubramaniam2010erratum} are used to describe interatomic interactions between Fe and H in the Fe-H system, while preventing unphysical H atom aggregation in bulk Fe. K-ART connects to the LAMMPS library's implementation of these potentials to obtain forces and energies~\cite{plimpton1995fast, sandiaLAMMPSMolecular}. While this potential introduces a shallow metastable state on the vacancy diffusion pathway~\cite{starikov2021angular}, other properties associated with defects are well reproduces~\cite{song2013atomic}.

\subsubsection{Simulation details}

Simulations are performed on a $7a_0 \times  7a_0 \times 7a_0$  Fe-atom cubic bulk crystal (686 atoms for the perfect crystal) with $a_{0}$ set to 2.8553\AA ~\cite{PhysRevB.84.214114}, the lattice constant for the bcc Fe crystal.
All systems are run at a KMC temperature of 300~K according to the transition state theory with events generated starting from energy minimized configurations. System boxes with H interstitials are first minimized using LAMMPS with the volume set to ensure $P = 0$ at 0 K. 

The total square displacement (SD) is computed as :
\begin{equation}\label{eq:7}
   SD = \sum_{i}^{N} {(r_i(t_n) - r_i(0))}^2,
\end{equation}
where $N$ is the number of particles and $r_i(t_n)$, the position of atom $i$ at KMC step $n$.
The ground state energy (GS) is defined in each simulation as that of the lowest energy minimum identified during the run. In the following, all energies are expressed with regard to the GS ($E(t_n) = E_{}^{\prime}(t_n) - E_{GS}$).

\subsubsection{Ab-initio ARTn}

We further validate the EAM results with DFT calculations using the ARTn-Quantum Espresso~\cite{giannozzi2009quantum,giannozzi2017advanced} (QE) package described in Ref.~\cite{jay2020finding}. For these calculations, a spin-polarized GGA-PBE exchange-correlation functional within the plane wave pseudopotential scheme as implemented in Quantum Espresso's (QE) software suite is used to relax the structure at the beginning. Calculations of activation energies and diffusion pathways are then performed using the ARTn coupled with QE code. To perform these calculations, we use a $4a_0 \times  4a_0 \times 4a_0$ 128 atoms with $a_{0}$ set to 2.834\AA~\cite{PhysRevB.91.104105}. The simulations use $\Gamma$-point calculations and are further tested for a $2\times 2 \times 2$ $k$-point mesh. Pseudopotentials are used to describe the interaction between core and valence electrons, with a kinetic energy cutoff of 40 Ry: $3 s^{2} 3 p^{6} 4 s^{2} 3 d^{10}$ and $1 s^{1}$, respectively, are treated as valence electrons in Fe and H. For this case, 0.01 eV/Å is the threshold value for total force for defining convergence at the saddle point.

\section{\label{sec:level1}Results}
\label{section:Results}
We first analyze the diffusion mechanisms for a single H atom moving in a perfect BCC iron matrix. Next, we  characterize the behavior of H atoms in the presence of vacancies and their effect on vacancy diffusion, focusing first on a mono-vacancy and, in the final subsection, a divacancy.

\subsubsection{Mono H Interstitial without Vacancy}

\begin{figure}[h]
   \centering
    \includegraphics[width = 9 cm,height = 5.5cm]{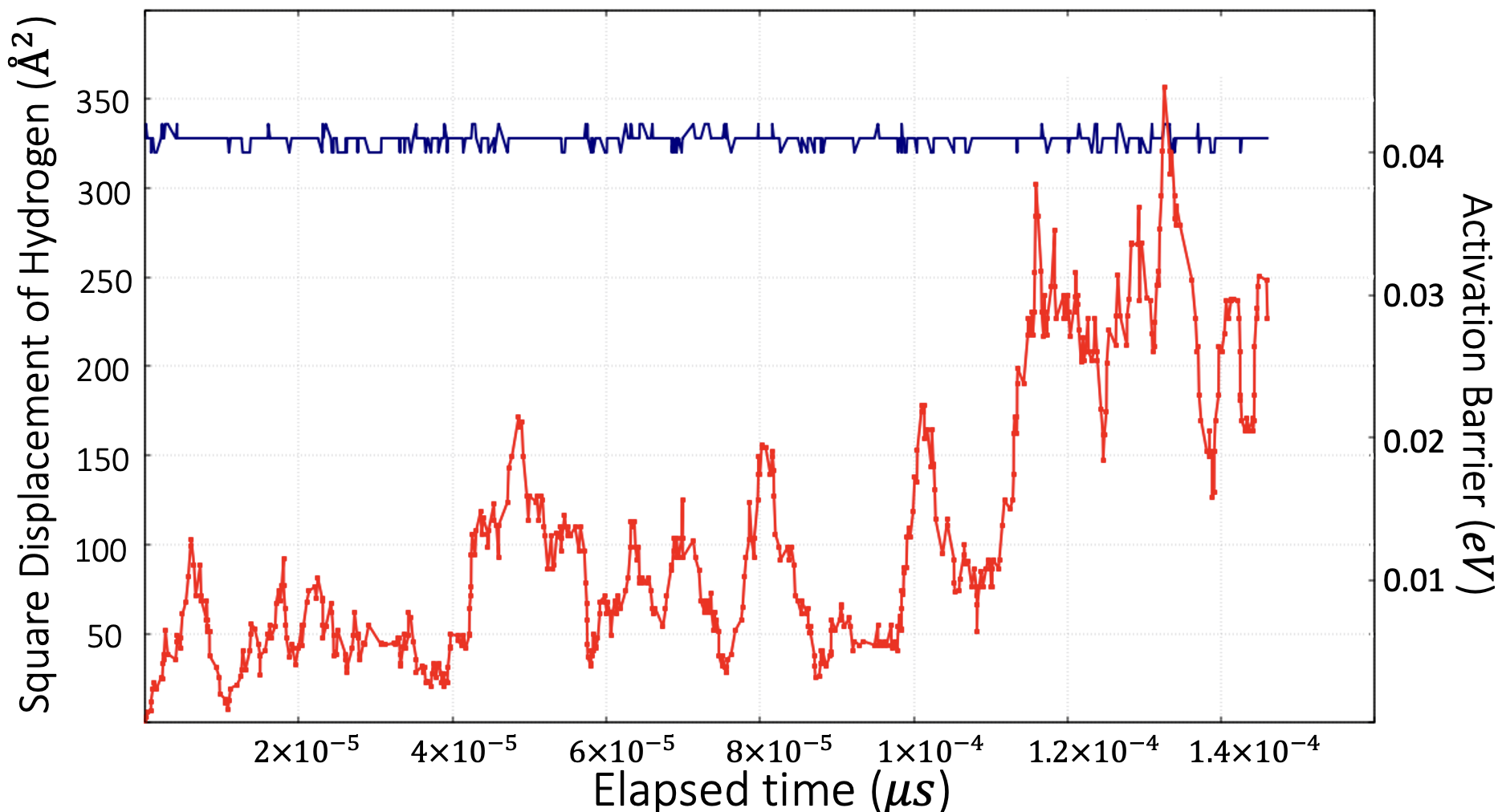}
    \caption{Squared displacement self-diffusion (left, red line) and activation barrier (right, blue line) as a function of time for a single H in a bcc crystal at 300 K, over 600 KMC steps.}
    \label{fig1}
\end{figure}

From an initial configuration consisting of a single isolated H atom placed in a tetrahedral site of the Fe-BCC, k-ART is launched for 600 KMC steps representing 0.14 ns of simulation time as shown in Fig~\ref{fig1}.

There are two high-symmetry interstitial sites in the bcc lattice: the tetrahedral and octahedral. As with previous computational studies on Fe, the ground state of the H mono interstitial is in tetrahedral sites for Fe BCC model~\cite{tateyama2003stability,jiang2004diffusion}. Experimental data also indicate that H occupies T sites primarily at low temperatures, while O-site occupancy becomes possible at high temperatures ~\cite{kiuchi1983effect,da1976solubility}.

Local lattice deformations around the interstitial H are very short range: 
an H atom added to the tetrahedral interstitial site results in a 0.179~\AA{} movement of the neighboring Fe atoms, which find themselves at 3.03~\AA{} of each other as compared to the equilibrium distance of 2.8553~\AA.
The H atom sitting in a T-site can hop to the next one along a curved pathway with a saddle point adjacent to an octahedral interstitial site. All four degenerate barriers are symmetrical, with a distance of 0.54~\AA\, from initial to saddle and 1.02~\AA\, from initial to final. The nearest-neighbor jumps require crossing a 0.04~eV activation barrier, consistent with DFT calculations~\cite{jiang2004diffusion, hayward2013interplay} which found 0.04~eV for the barrier and experiments~\cite{nagano1982hydrogen} findings the barrier 0.035 (240-970 K).  
The results of our study are in good agreement with those of other studies~\cite{wen2021new,ramasubramaniam2010erratum}, so we can move forward with the assessment of structures with deficiencies, such as vacancies. This event's prefactor, obtained using hTST,  is found to be $2.22 \times 10^{13}$~Hz with a transition event rate of $4.49 \times 10^{12}$~Hz.

\subsubsection{Trapping of H atoms inside the vacancy}

\begin{figure*}
\begin{minipage}[h]{0.47\linewidth}
\begin{center}
\includegraphics[width=1\linewidth]{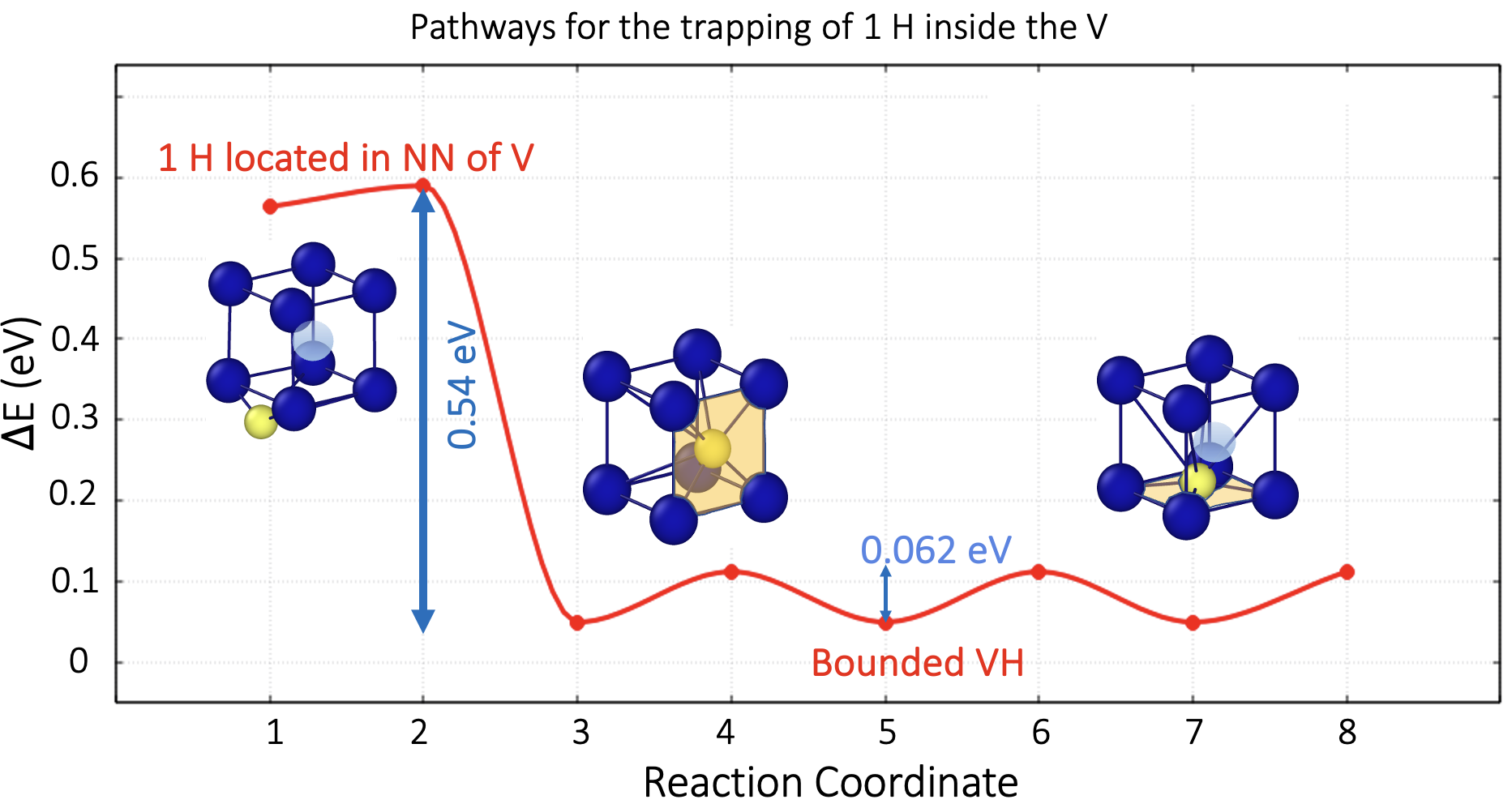} 
\label{a}
\caption*{(a)}
\end{center}
\end{minipage}
\hfill
\vspace{0.1 cm}
\begin{minipage}[h]{0.47\linewidth}
\begin{center}
\includegraphics[width=1\linewidth]{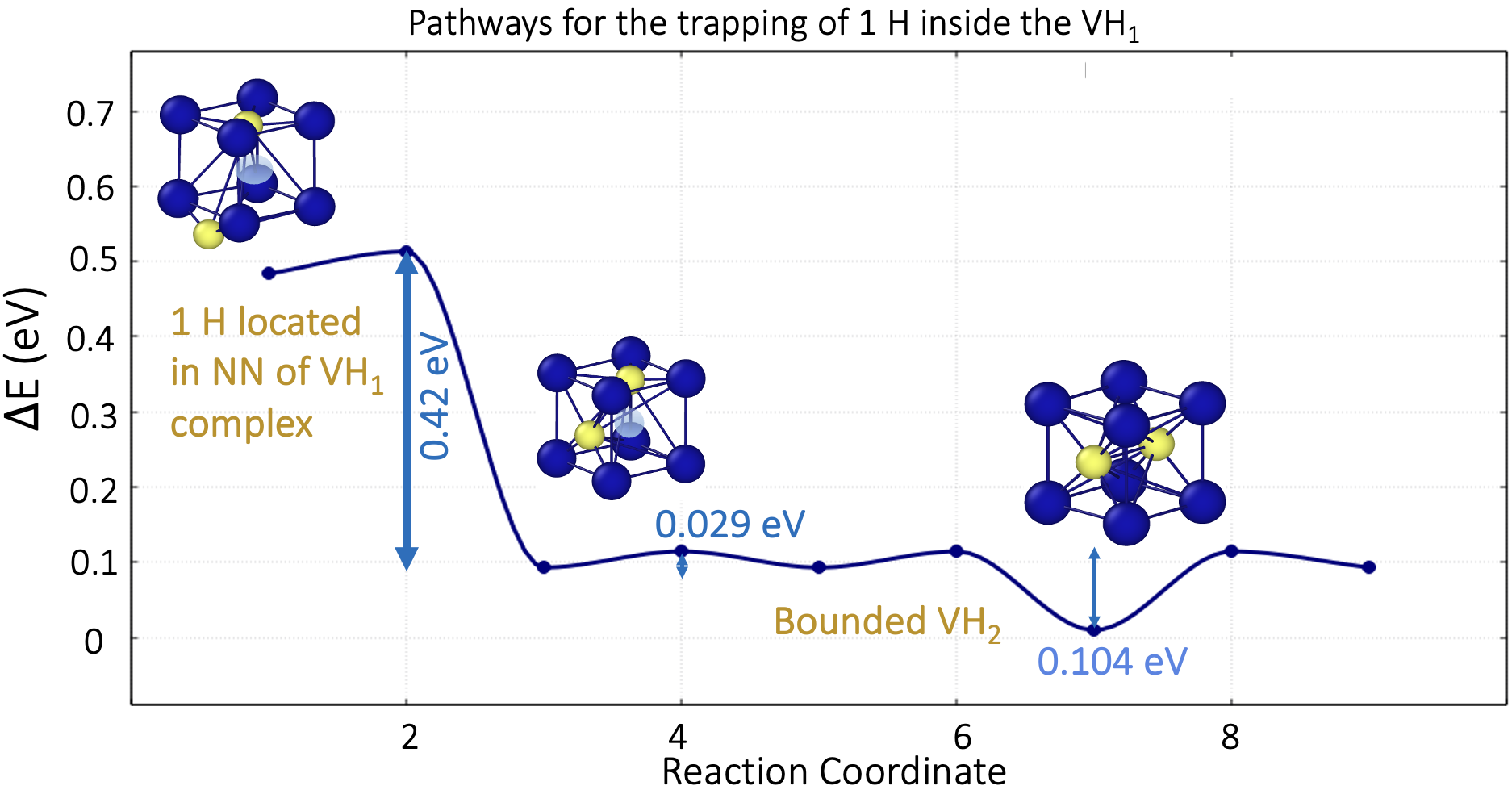} 
\label{b}
\caption*{(b)}
\end{center}
\end{minipage}
\vfill
\vspace{0.2 cm}
\begin{minipage}[h]{0.47\linewidth}
\begin{center}
\includegraphics[width=1\linewidth]{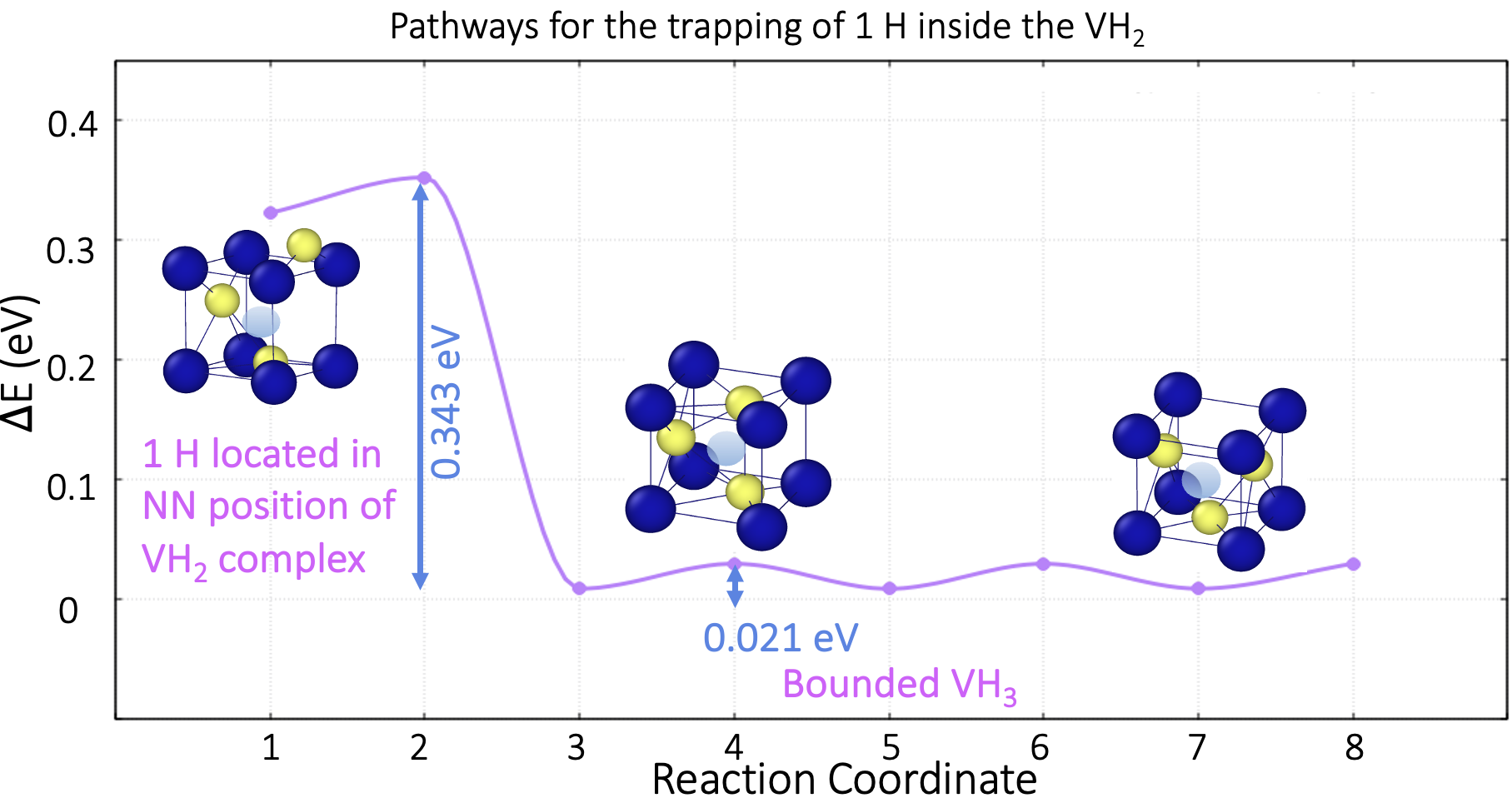} 
\label{c}
\caption*{(c)}
\end{center}
\end{minipage}
\hfill
\begin{minipage}[h]{0.47\linewidth}
\begin{center}
\includegraphics[width=1\linewidth]{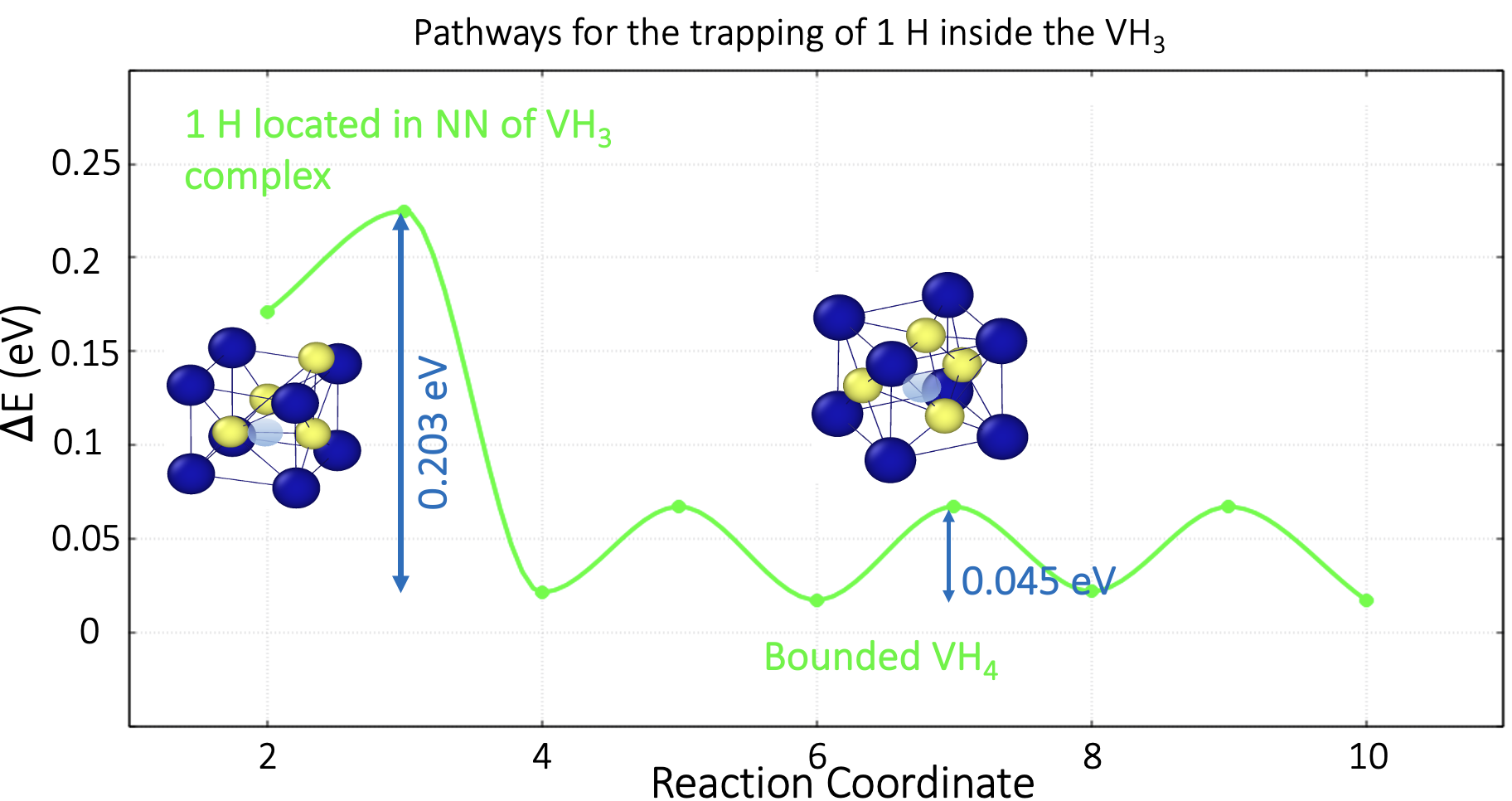} 
\label{d}
\caption*{(d)}
\end{center}
\end{minipage}
\caption{Energy diffusion pathways associated with H trapping into a vacancy (VH$_x$). (a) Starting with an empty vacancy, trapping of a first H atom; (b) adding a second H atom to VH$_1$; (c) adding a third H to VH$_2$ and, (d), a fourth H to VH$_3$.
Inserts: A blue color indicates an iron atom, while a yellow color indicates a H atom. At the center, there is a vacancy indicated by a gray-blue color.}
\label{fig2}
\end{figure*}

We now look at the trapping of H atoms inside the vacancy. Elastic deformations caused by the defect are small enough that diffusion barriers for an H atom placed in the second closest tetrahedral interstitial site of the vacancy are not affected at any noticeable level by the vacancy. To focus on H and vacancy interactions, k-ART simulations are therefore launched  with the H positioned in a  tetrahedral interstitial site at the first-nearest neighbor of the vacancy. From this position, the H atom can diffuse to the nearest tetrahedral interstitial sites, away from the vacancy, crossing ~0.05-0.07 eV barriers, depending on the diffusion direction, or diffuse into the vacancy crossing an energy barrier of 0.025~eV, where trapping occurs with an energy barrier of 0.54~eV (see Fig.~\ref{fig2}(a)). 

Once trapped within the vacancy, the H ground state is located on the 0.255 \AA{} offset of an octahedral site within the vacancy, as observed previously~\cite{mirzaev2014hydrogen}. It can move to one of the four neighboring similar sites by crossing a 0.06~eV barrier. On this path, the H atom travels 0.82 \AA\ from the initial to the saddle point for a total movement, after following a symmetric path, of 1.60 \AA\ from the initial to the final point. The H can also leave the vacancy, crossing a 0.54~eV barrier. The computed hTST prefactor for an H atom to jump over a 0.026~eV barrier to enter the vacancy is equal to $4.5 \times 10^{13}$~Hz.
For the second event with a barrier of 0.06~eV, related to the H barrier to diffuse within the vacancy, jumps from one offset of an octahedral to the nearest offset of an octahedral interstitial site inside the vacancy occur with a transition event rate of $8.87 \times 10^{13}$~Hz, associated with computed hTST prefactor of $1.87 \times 10^{13}$~Hz. Our results show that, contrary to other complex systems such as high entropy metallic alloys~\cite{sauve2022unexpected}, the H diffusion prefactor is not significantly affected by its insertion into a vacancy.

\begin{figure}[h]
   \centering
    \includegraphics[width = 8.5cm,height = 5.5cm]{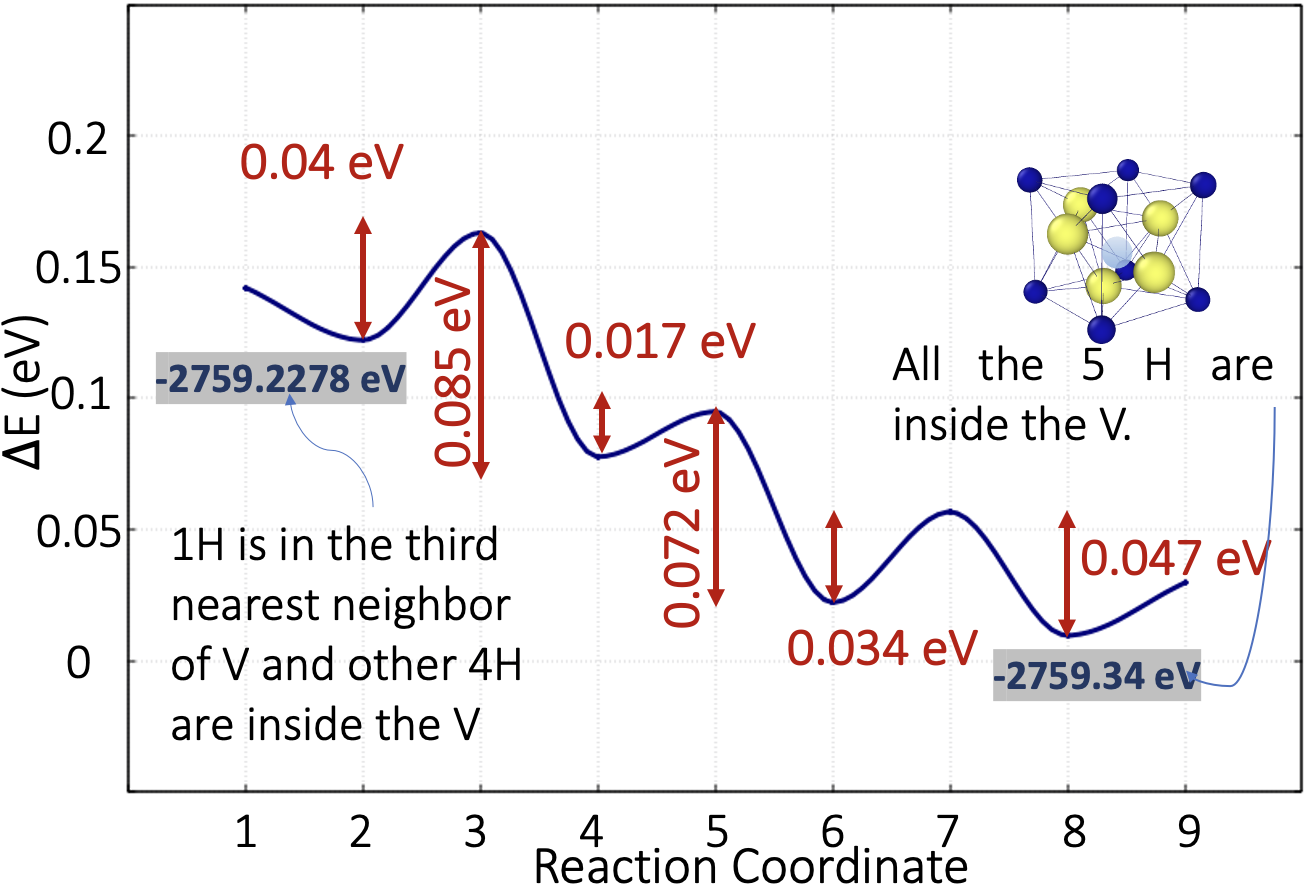}
    \caption{Energy diffusion pathway associated with trapping a 5th H into a vacancy containing already four H atoms (VH$_4$). The inset is the same as in \Cref{fig2}.}
    \label{fig3}
\end{figure}

With a first H trapped in the vacancy, we add a second one on the tetrahedral nearest neighbor of the vacancy as shown in Fig~\ref{fig2}(b), in order to characterize the evolution of the energy landscape surrounding the vacancy as a function of the number of bound H atoms. In each case, we show the H insertion pathways into the vacancy as H atoms are added one by one onto the vacancy's nearest neighbor. The top left (a) diffusion path in Fig~\ref{fig2} presents the insertion of a single H atom in an empty vacancy.  Fig~\ref{fig2}(b) starts from this trapped configuration and shows the energy landscape associated with a second H atom coming into the vacancy. This sequence is pursued with a third (c) and a fourth H atom (d). While up to five H atoms can be inserted into the vacancy, the addition of a sixth atom requires the detrapping of one H atom first. Adding more H leads to a lower barrier for H atom detrapping from the vacancy, starting with 0.54~eV for the first trapped H, 0.42~eV for the second, 0.34~eV for 3 H atoms, 0.203~eV for 4 H and 0.046~eV for 5H.  In our simulation, the barrier for the 5th H atom to trap inside the vacancy is 0.034~eV, while the reverse barrier is 0.047~eV, close to the barrier of the H in bulk, when no vacancy exists. Any attempt to add a 6th H to the vacancy leads to a barrier-less detrapping. Moreover, while the elastic deformation around the vacancy is small with a single trapped H, these increase as more H are inserted.  Fig~\ref{fig3} shows the energy landscape as a 5th H atom gets trapped inside the VH$_4$ complex.

Table~\ref{tab:table1} provides the incremental binding energies, discussed in the method section, comparing with other EAM~\cite{restrepo2020effect} and DFT without and with zero-point energy corrections~\cite{tateyama2003stability, hayward2013interplay} studies.  Our results are within an accuracy range of other numerical studies (see Table~\ref{tab:table1}). In perfect BCC iron, the H solution energy is calculated as 0.29~eV using \Cref{eq:solution} (where the $E_{H_2}$ equals -4.738~eV), consistent with previous studies~\cite{PhysRevB.82.235125}. Binding energies are also in agreement with thermal desorption spectroscopy results~\cite{iwamoto1999superabundant} where 1-2 H atoms are estimated to have a binding energy of 0.632~eV, while 3-6 H atoms have a binding energy of 0.424~eV. We note that the difference between the energies shown in the Fig~\ref{fig2} and the incremental binding energy (\Cref{tab:table1}) is due to the fact that the latter is obtained using \Cref{eq:6}.

\begin{table}[h]
\caption{\label{tab:table1}%
Incremental binding energy of the \textit{x}th H atom inserted into a monovacancy (in eV) for this (first column) and previously published work. Column 2 presents EAM results from Restrepo \textit{et al.}~\cite{restrepo2020effect}. Columns 3 and 4 present results from DFT calculations from Tateyama \textit{et al.}~\cite{tateyama2003stability} and Hayward \textit{et al.}~\cite{hayward2013interplay} without zero-point energy corrections. Column 5 shows Hayward \textit{et al.}~\cite{hayward2013interplay}'s results with ZPE.}
\begin{ruledtabular}
\begin{tabular}{cccccc}
\textrm{x}&
\textrm{$E_B$}&
\textrm{EAM~\cite{restrepo2020effect}}&
\textrm{DFT~\cite{tateyama2003stability}}&
\textrm{DFT~\cite{hayward2013interplay}}&
\textrm{DFT(ZPE)~\cite{hayward2013interplay}}
\\
\colrule
1   &   0.603   &   0.603   & 0.559 & 0.498 & 0.616\\
2 & 0.561 & 0.552 & 0.612 & 0.543 & 0.651\\
3 & 0.322 & 0.298 & 0.399 & 0.337 & 0.381 \\
4 & 0.213 & 0.182 & 0.276 & 0.304 & 0.351\\
5 & 0.0795 & 0.056 & 0.335  & 0.269 & 0.296\\
\end{tabular}
\end{ruledtabular}
\end{table}

\subsubsection{Effect of H on vacancy diffusion}

The previous section describes the trapping of H atoms inside a vacancy. We now explore the effect of the presence of H atoms on the kinetics of an isolated Fe vacancy. As (i) H diffuses faster through the interstitial network than the vacancy and (ii) the barrier to H detrapping is lower than the vacancy (V) diffusion barrier, we focus here only on the bound complexes (VH$_x$, where $x$ indicates the number of bound H).

\begin{figure}[b]
    \centering
    \includegraphics[height = 5.5cm, width = 8.5 cm]{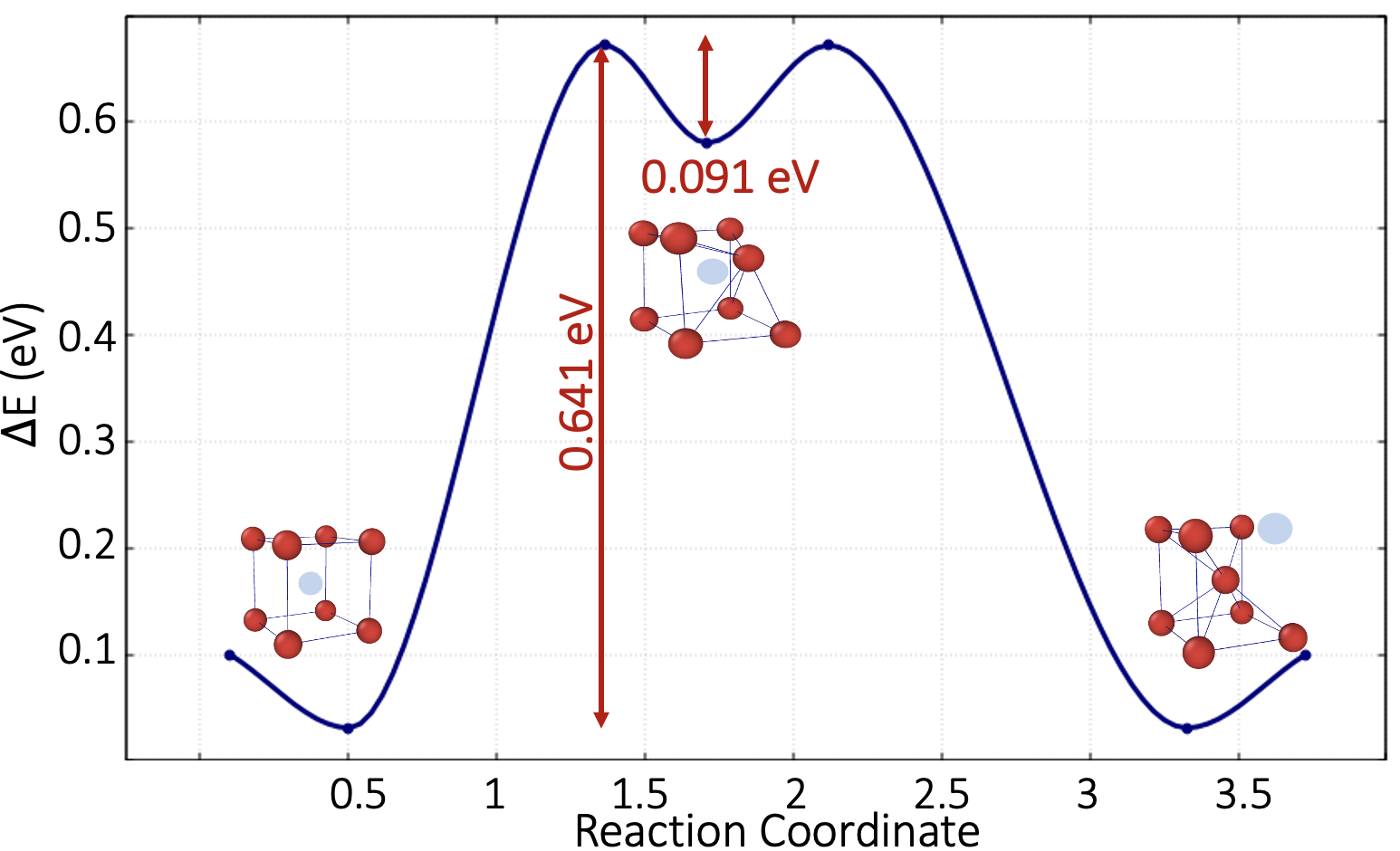}
    \caption{Diffusion pathway for an isolated vacancy in a perfect Fe bcc crystal.}
    \label{fig4}
\end{figure}

In isolation, the vacancy diffuses to one of the eight first-neighbor positions through a two-step process (Fig~\ref{fig4}): moving over a 0.641~eV barrier, the vacancy moves into a metastable intermediate interstitial state at 0.55~eV above ground state~\cite{restrepo2016diffusion}. At this point, the iron atom can take two symmetrical paths over equal 0.091~eV barriers that either bring the vacancy back to its initial position or move it into a neighboring position,  resulting in a first-neighbor vacancy hop. 

\begin{figure}[h]
    \centering
    \includegraphics[width = 4cm]{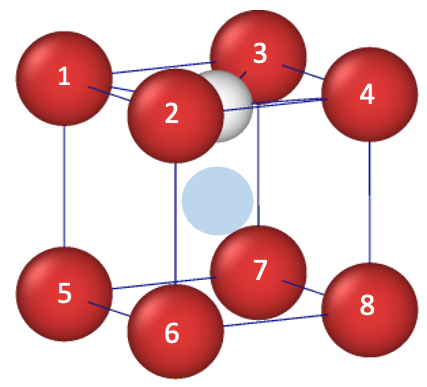}
    \caption{Fe atoms surrounding a vacancy numbered with respect to a trapped H position. Red indicates Fe atoms, white the H and gray-blue, the vacancy site. Fe labels are used in Table~\ref{tab:table2}.}
    \label{fig5}
\end{figure}

By inserting one H atom into the vacancy (VH$_1$), the energy landscape associated with vacancy diffusion pathways becomes more complex as the eight adjacent iron atoms are now in unequal environments. Our simulations show that the barrier for iron atoms in the nearest neighbor of the vacancy to diffuse into the vacancy depends on the H atom's position. Fig.~\ref{fig5} shows the position of Fe atoms adjacent to the vacancy in relation to the trapped H.  Table~\ref{tab:table2} lists the energy barriers for adjacent Fe atoms diffusing into the vacancy. There are three different mechanisms in which the Fe atoms marked one to four in Fig.~\ref{fig5} can diffuse, crossing 0.726, 1.105, and 1.39~eV barriers, respectively. Associated pathways lead the vacancy to reach a first-neighbor site in either one (1.39~eV barrier) or two steps (barriers of 0.726 and 1.105~eV). Looking more specifically at the barriers that are shown in table~\ref{tab:table2}, we see the inverse barrier for 1.39~eV mechanism is 0.805~eV, which indicates that the one-step mechanism is not symmetric and does not return the system to the same minimum potential as the table only presents the first barriers but not the complete mechanisms.  It should be noted, moreover, that the barriers for Fe to move into the vacancy are calculated with an H in a fixed position. As shown in Fig.~\ref{fig2}, H moves quickly among sites within the vacancy, so the analysis presented here still means that VH$_1$ diffusion is essentially isotropic in crystal, irrespective of the H position in the vacancy.

\begin{table}[h]
\caption{\label{tab:table2}
List of the barriers for first-neighbor Fe atoms to move into the vacancy occupied with $H$ atom (VH$_1$). Fe atoms are numbered according to their position with respect the H (see \Cref{fig5}).
Rates are computed using \Cref{eq:4}. Root-mean square displacement $d_{si}$ ($d_{fi}$) between the saddle point (final minimum) and the initial state are also indicated.
}
\begin{ruledtabular}
\begin{tabular}{cccccc}
\scriptsize{ \begin{tabular}{@{}c@{}}Atom \\ ID\end{tabular}} &\scriptsize{ \begin{tabular}{@{}c@{}}Barrier \\ (eV)\end{tabular}}& \scriptsize{ \begin{tabular}{@{}c@{}}Inverse \\ Barrier (eV)\end{tabular}}& $\Gamma_{if}$ (Hz) & $d_{si}$ 
(\AA) & $d_{fi}$ (\AA)\\
\hline
\\
Fe 1-4 & 1.105 & 0.591 & 2.65$\times 10^{-6}$& 1.767  &  2.521 \\
\\
 & 1.391 & 0.805 & 4.23$\times 10^{-11}$ & 1.657  &  2.653 \\
\\
 & 0.726 & 0.14 & 6.23 & 1.215  &  1.913 \\
\\
 Fe 5-8 & 0.679 & 0.0932 & 3.83$\times 10^{1}$ & 0.8813 & 1.224 \\
\end{tabular}
\end{ruledtabular}
\end{table}

\begin{figure}[h]
    \centering
    \includegraphics[width = 8.5cm]{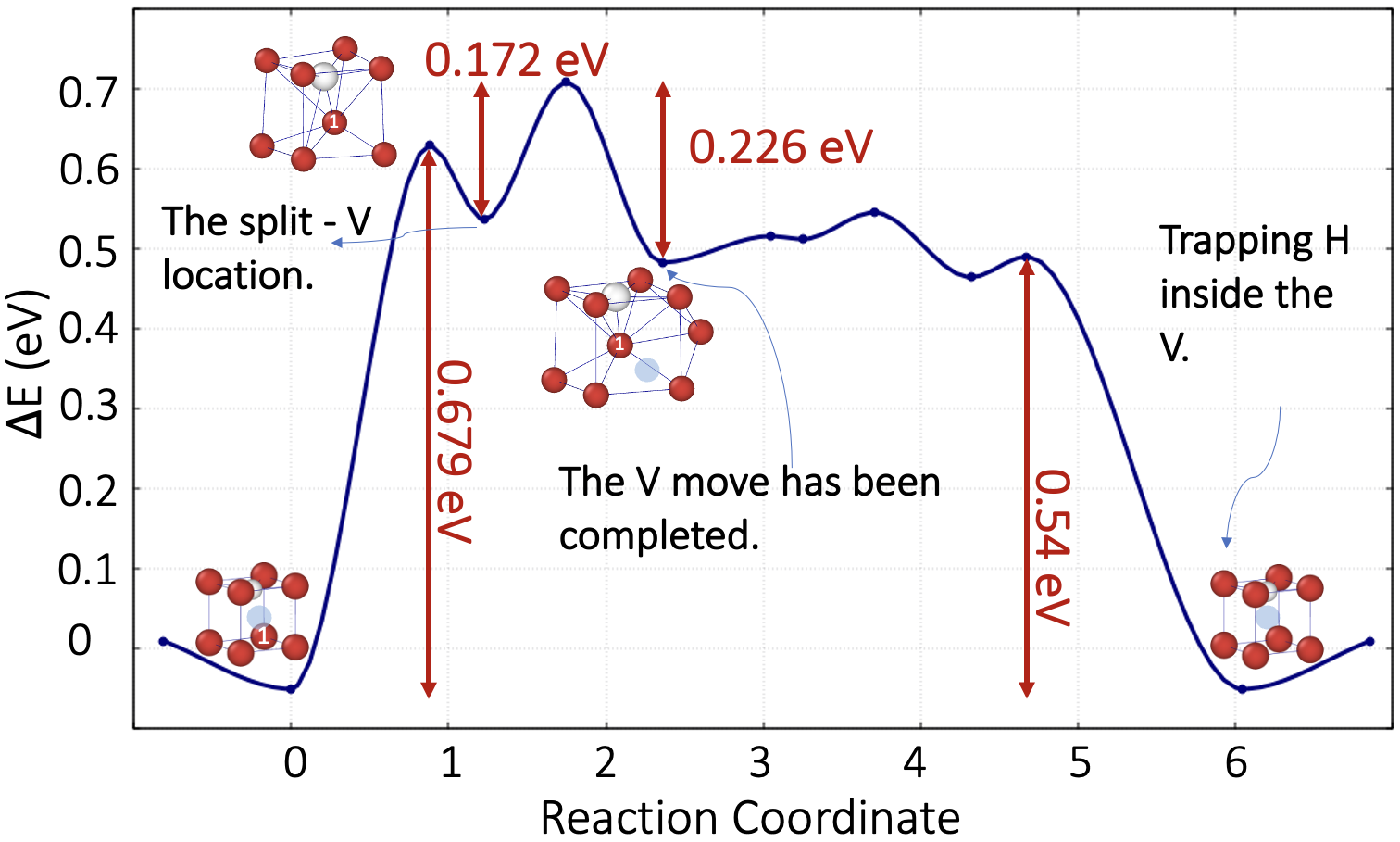}
    \caption{Diffusion path for the VH$_1$ complex. A Fe jumps a 0.68~eV to reach metastable split-vacancy position in the first step. The vacancy completes its move crossing a 0.17~eV barrier.  Left behind the vacancy, the H atom follows it in three steps, crossing low energy barriers.}
    \label{fig6}
\end{figure}

The lowest insertion barrier for Fe neighboring a vacancy with a trapped H is at 0.679~eV, slightly above the 0.641~eV barrier associated with the diffusion of an isolated vacancy. It is available to atoms marked five to eight in Fig.~\ref{fig5}. This 0.679~V barrier is the first of a multiple-step pathway that brings the VH$_1$ complex into a neighboring site. As shown in Fig.~\ref{fig6}, crossing the 0.679~eV barrier (the moved atom is represented by ``1'' in Fig.~\ref{fig6}) brings a Fe atom into a metastable position (0.58~eV above minimum) that creates a EAM-characteristic split vacancy with the H atom in its original place. Moving over a  0.172~eV barrier completes the vacancy move: we now have an empty vacancy site moved by one lattice spacing and an interstitial H into a near metastable octahedral state. As the H atom is left behind the vacancy, another step, with a 0.02~eV barrier (0.54~eV inverse barrier) as shown in Fig.~\ref{fig6} finishes the move of the H atom and brings the VH$_1$ into a new lattice site. The three steps shown as 2-5 along the reaction coordinate in Fig.~\ref{fig6} are associated with H diffusion, as it follows and reintegrates the vacancy. Since the initial H position is off the octahedral site, the H atom crosses small barriers to reach a tetrahedral interstitial site and, from there, moves into the vacancy in the same manner as described earlier. Overall, following the lowest-energy path, the VH$_1$ complex diffusion requires crossing an effective barrier of 0.758~eV that defines the mechanism as concerted, which is 0.117~eV higher than for an isolated vacancy and is 0.346~eV lower than the next lowest diffusion mechanism. Note that once the vacancy has moved, it is also very likely for the $H$ to diffuse away from the vacancy, breaking the VH complex.

First de-trapping the H opens the door to multiple pathways with energy higher than the concerted mechanism. In one of our simulations, for example, where first the H atom de-traps from the vacancy, followed by the vacancy diffusion, we observe a total effective barrier of 1.104~eV, higher than the concerted motion, with the single H first crossing a 0.54~eV barrier, followed by the diffusion of the isolated vacancy with a 0.67~eV barrier. The concerted mechanism is therefore favored for the diffusion of the bound VH configuration. 

\begin{figure}[h]
    \centering
    \includegraphics[width = 8.5cm]{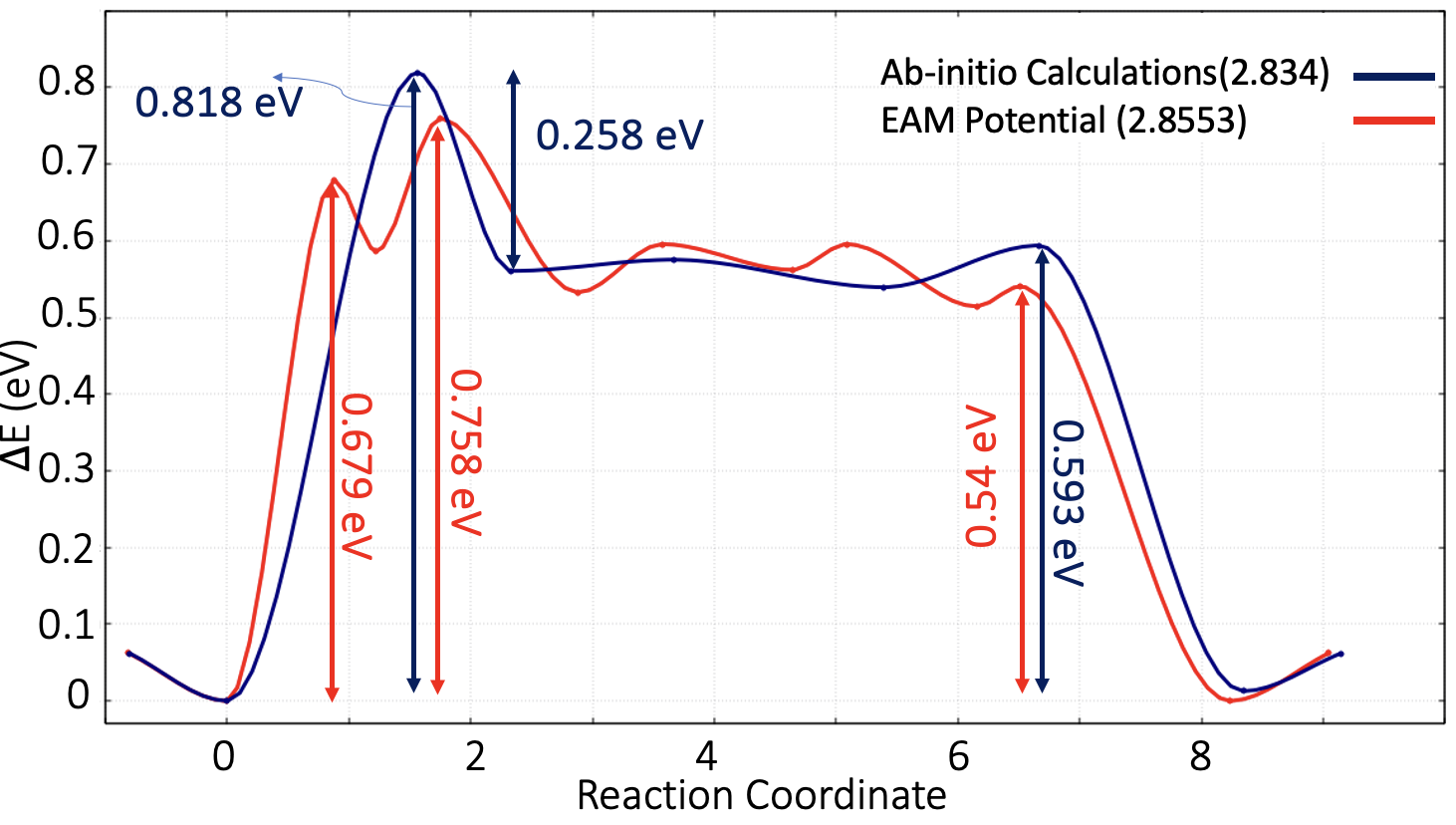}
    \caption{Comparison of the diffusion path for the VH$_1$ complex obtained from k-ART-EAM (red) and ab initio-ARTn (navy) calculations.}
    \label{fig7}
\end{figure}

To assess the validity of this event, we perform ab initio-ARTn simulations for a system with 127 Fe plus 1 H atoms with a VH$_1$ complex. As with the kART+EAM simulation, the ground state position of the H atom in the vacancy is in an offset of the octahedral interstitial sites. Fig.~\ref{fig7} shows the diffusion pathway of the VH$_1$. As a result of this calculation, the barrier for VH$_1$ complex diffusion is increased to 0.818~eV with $\Gamma$-point calculations and 0.782~eV using $2\times 2 \times 2$ k-point mesh, which is comparable to Ref.~\cite{hayward2013interplay}'s DFT calculations. Fig. ~\ref{fig7} shows the comparison between the DFT and associated empirical pathway showing great similarity except for the spurious shallow metastable minimum at the saddle-point, a known artifact with these empirical potentials~\cite{starikov2021angular} as mentioned earlier in methodology. This supports the EAM's potential to look at more complex defects within large simulation boxes, which are too costly to evaluate with DFT.

Turning the vacancy with two H complexes (VH$_2$), we study two classes of diffusion similarly to the VH$_1$ complex. Diffusion can take place with the vacancy dragging the H atoms (VH$_2$) or with one of the H atoms first unbinding (VH$_1$,1H), letting the VH$_1$ jump to a nearby site and reinserting the vacancy (VH$_2$). For the first class of trajectory, we find three different types of mechanisms depending on where the H atoms are located inside the vacancy, with energy barriers of 0.755, 1.15, and 1.65~eV at the first step of the mechanism. As with the single H, these barriers are the first steps for longer pathways allowing the VH$_2$ complex to move to a nearest-neighbor site.

\begin{figure}[h]
    \centering
    \includegraphics[width = 8.4cm]{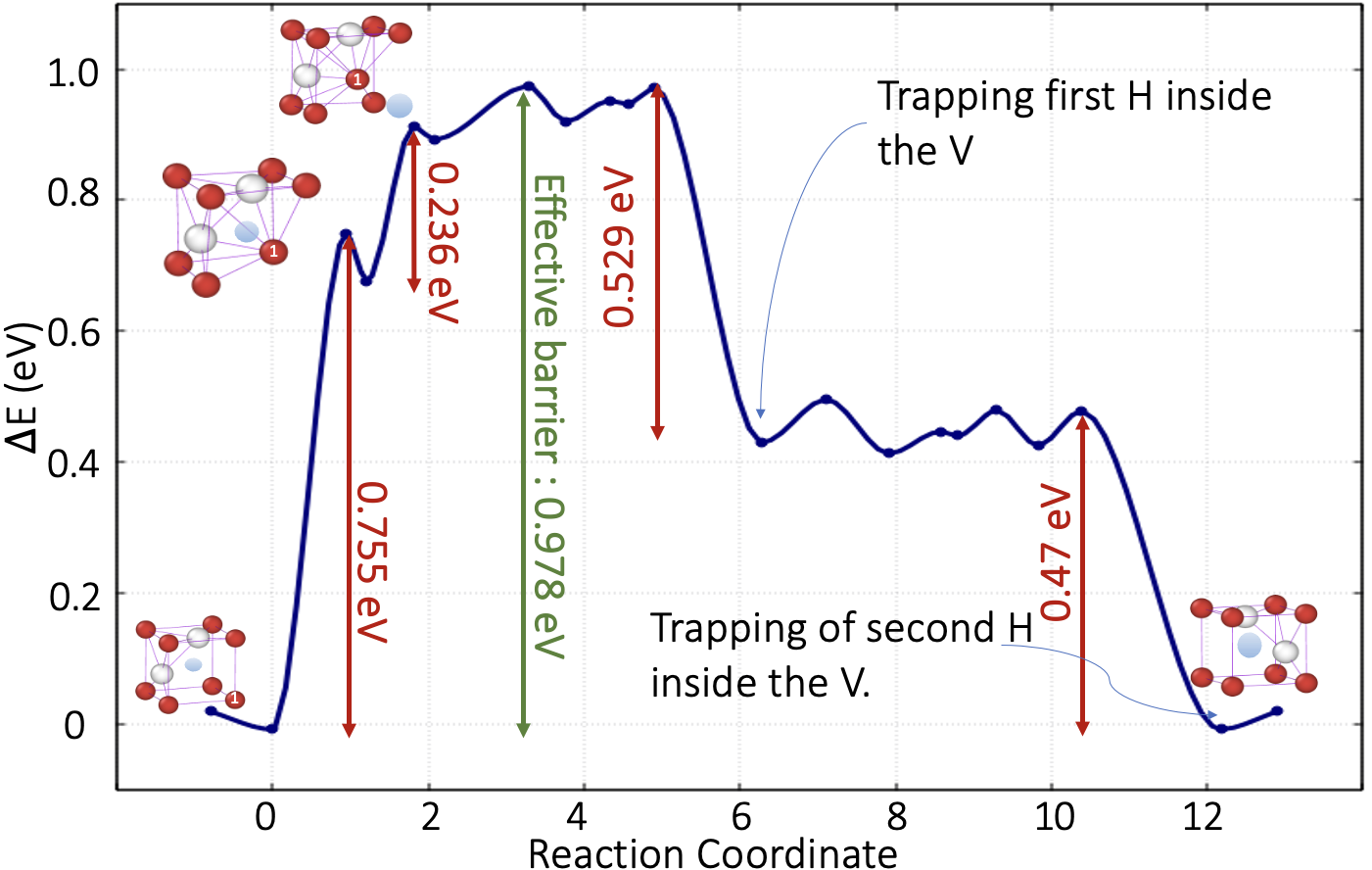}
    \caption{Diffusion path for the VH$_2$ complex. A Fe (represented by ``1'') atom crosses a 0.75~eV barrier to reach an EAM-characteristic split-vacancy configuration. The second step, with a 0.236~eV barrier, completes the move of the vacancy into its next neighbor crystalline site. At that point, both H are outside of the vacancy, with the configuration sitting 0.9~eV above the minimum-energy configuration. The first H moves into the vacancy in three steps, with a total barrier of 0.078~eV, leading to a 0.52~eV relaxation as the first H atom jumps into the vacancy. The second H follows in four steps, with a barrier of 0.05~eV, providing an additional 0.47~eV relaxation.}
    \label{fig8}
\end{figure}

Diffusion barriers for the vacancy are strongly affected by the H position within the vacancy, as previously discussed. The most probable mechanism, with a 0.755~eV barrier, is associated with the two H positioned on neighboring offset octahedral sites aligned in the (111) direction (Fig~\ref{fig8}). The Fe atom (represented by ``1'' in Fig~\ref{fig8}) crosses a 0.75~eV barrier to reach the EAM-characteristic split-vacancy configuration. By the second step, associated with a barrier of 0.236~eV, the vacancy has moved into its neighboring crystalline site. In this configuration, both H are outside of the vacancy, and the configuration sits 0.9~eV above the minimum-energy configuration. The first H moves into the vacancy in three steps, with a total barrier of 0.078~eV, leading to a 0.52~eV relaxation as the first H atom jumps into the vacancy. The second H follows in four steps, with a barrier of 0.05~eV, providing an additional 0.47~eV relaxation. This class of diffusion has an effective energy barrier of around 0.978~eV. Clearly, here also, once the vacancy has moved, H can diffuse away with high probability.

\begin{figure}[h]
    \centering
    \includegraphics[width = 8.3cm, height = 5.5 cm]{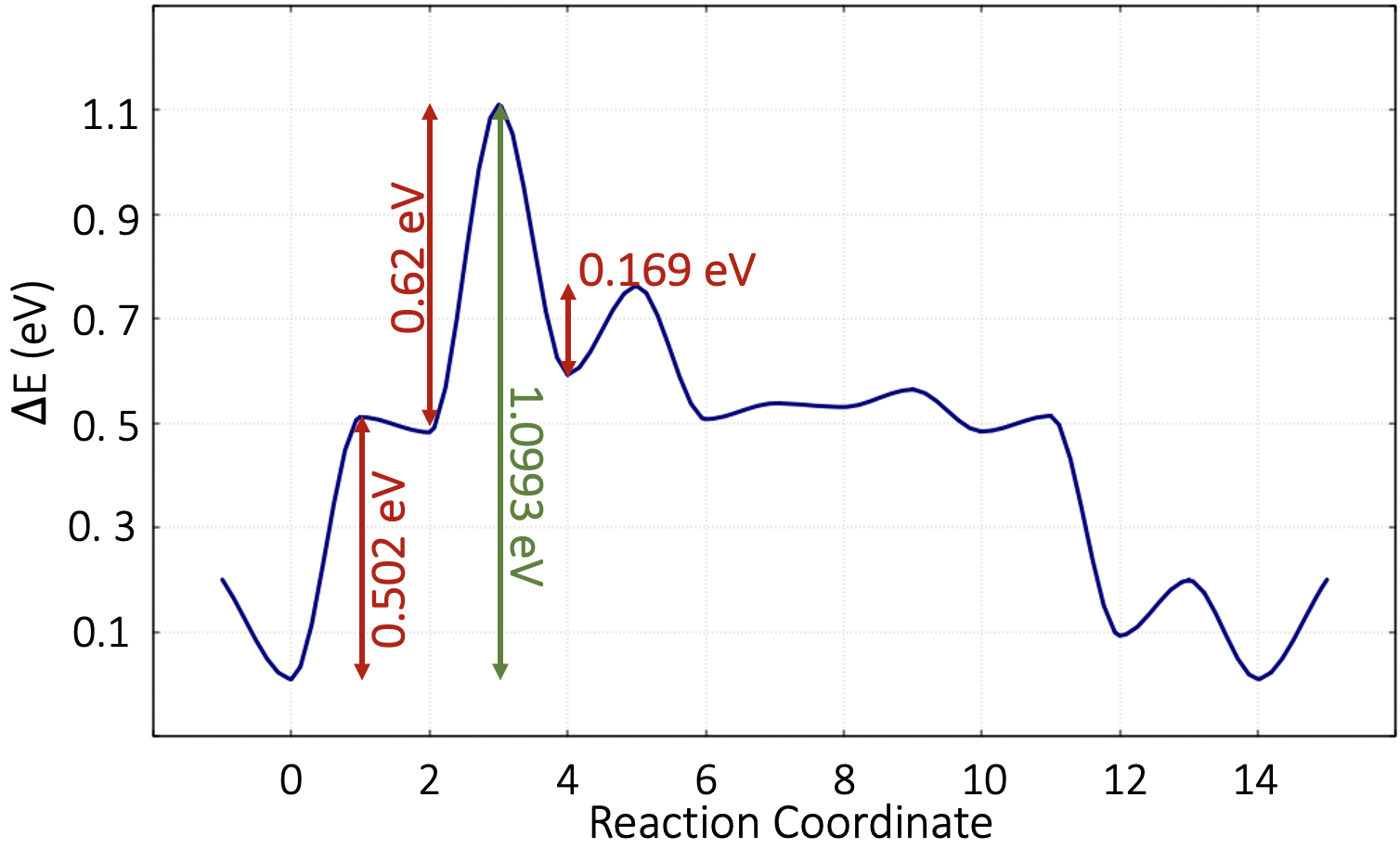}
    \caption{The diffusion pathway for the VH$_2$ complex when one of the H atoms first detraps from the vacancy, crossing a 0.502~eV barrier which results in VH$_1$ + 1H configuration. As for the VH$_1$ motion, a Fe then moves over a 0.620~eV barrier into a EAM-characteristic split vacancy metastable configuration. The Fe completes its motion, moving the vacancy by one site after crossing a 0.169~eV barrier and forming a new VH$_1$ + 1H state. During steps 8-14, the H left behind the vacancy diffuses into the displaced vacancy.}
    \label{fig9}
\end{figure}

The second class of diffusion mechanisms for the VH$_2$ complex involves first a detrapping of H. Restrepo, Lambert and Paxton~\cite{restrepo2020effect} suggested that the dominant mechanism requires first for one of the two H atoms to move out of the vacancy. With one of the H atoms detrapped, a Fe atom adjacent to the vacancy moves into the empty site. Hence, we have an H atom inside the vacancy, with the other H atom lying outside of the vacancy in the first-nearest neighbor. Their research refers to this additional trapped H atom as a ``helper'' atom since it facilitates the migration of the vacancy. Our simulations show that the H atom first detraps from the VH$_2$ with a 0.502~eV barrier, followed by a Fe jump into an EAM-characteristic split-vacancy site with a barrier of 0.620~eV. A second step in the diffusion of Fe has an energy barrier of 0.169~eV and brings the Fe from the previous position to the vacancy position (a total energy barrier of 1.099~eV). In this configuration, we have one H inside the vacancy and the other left behind the vacancy site, which means that to complete the diffusion mechanism, the second H must come and trap inside the vacancy (see Fig.~\ref{fig9}). While we observe, as previously, a two-step motion for the vacancy diffusing, with a 0.620~eV barrier into a split vacancy followed by a 0.169~eV barrier to finalize the motion, Ref.~\cite{restrepo2020effect} find a single 0.289~eV barrier. This difference, which is mostly likely due to a step missing by Restrepo \textit{et al.}, is analyzed in Sect. \ref{section:Discussion}. 

Overall, therefore, we find that, for VH$_2$, both vacancy diffusion first (0.978~eV) and a mechanism where a H detraps first (a total energy barrier of 1.099~eV) are competing mechanisms with very similar probability, contrary to previous findings~\cite{restrepo2020effect}.

\begin{figure}[h]
    \centering
    \includegraphics[width =  8cm, height = 6cm]{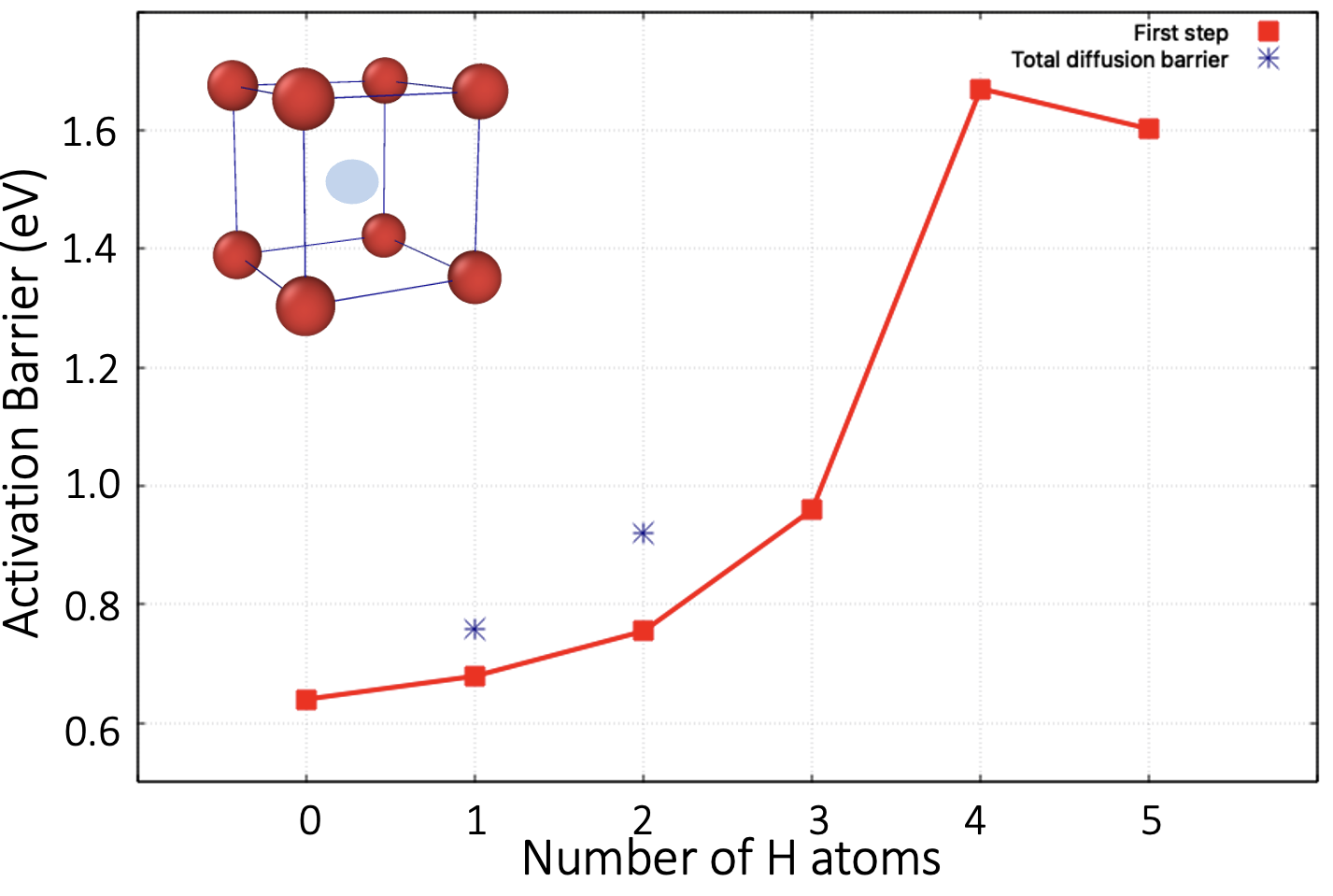}
    \caption{Lowest energy barrier (most probable barrier) for one of the 8 iron atoms next to the vacancy to move into a split interstitial, first step towards moving the vacancy, as a function of the number of H ($x$) trapped into the vacancy (VH$_x$). The lowest total barrier for moving the vacancy (leaving some or all the H behind) is shown for $x=1$ and 2.}
    \label{fig10}
\end{figure}

As more H are trapped in the vacancies, the number of potential pathways for the VH$_x$ complex increases. In all cases, however, diffusion of the vacancy becomes more and more difficult as little space is available for a Fe atom to diffuse into the vacancy. A list of all possible first barriers to have diffusion is listed in~\Cref{tab:table3} as a function of the number of H atoms. For example, when we have 4 and 5 H atoms, there is only one barrier for vacancy jumping, excluding the first detrapping of an H. We note, however, that, in some cases, the motion of the Fe can force detrapping of a H. Fig.~\ref{fig10} shows the lowest barrier for the first step in the diffusion of a Fe in the vacancy. The activation barrier for vacancy diffusion increases systematically with the number of trapped H, as space is less available for movement. 

\begin{table}[!]
\caption{\label{tab:table3}
Energy barriers for the first Fe diffusion step towards the vacancy (to form a split vacancy) as a function of the number of trapped H atoms. }
\begin{ruledtabular}
\begin{tabular}{c|cccccc}
\scriptsize{\begin{tabular}{@{}c@{}}Number of \\ H atoms\end{tabular}} & \multicolumn{6}{c}{\scriptsize{Barriers (eV)}} \\
\hline

1 & 0.679 & 0.726 & 1.105 &1.39\\
2 & 0.75 & 0.84 & 1.15 & 1.24 & 1.65 & 1.73\footnote{This barrier is for a configuration where a H atom first detraps.}\\
3 & 0.963 \footnotemark[\value{mpfootnote}]& 1.33 \footnotemark[\value{mpfootnote}]
& 1.54 \\
4 & 1.675\\
5 & 1.602 \footnotemark[\value{mpfootnote}]\\
\end{tabular}
\end{ruledtabular}
\end{table}

\subsubsection{Effect of H atoms on divacancy diffusion}

\begin{figure}
    \centering
    \includegraphics[width = 8cm]{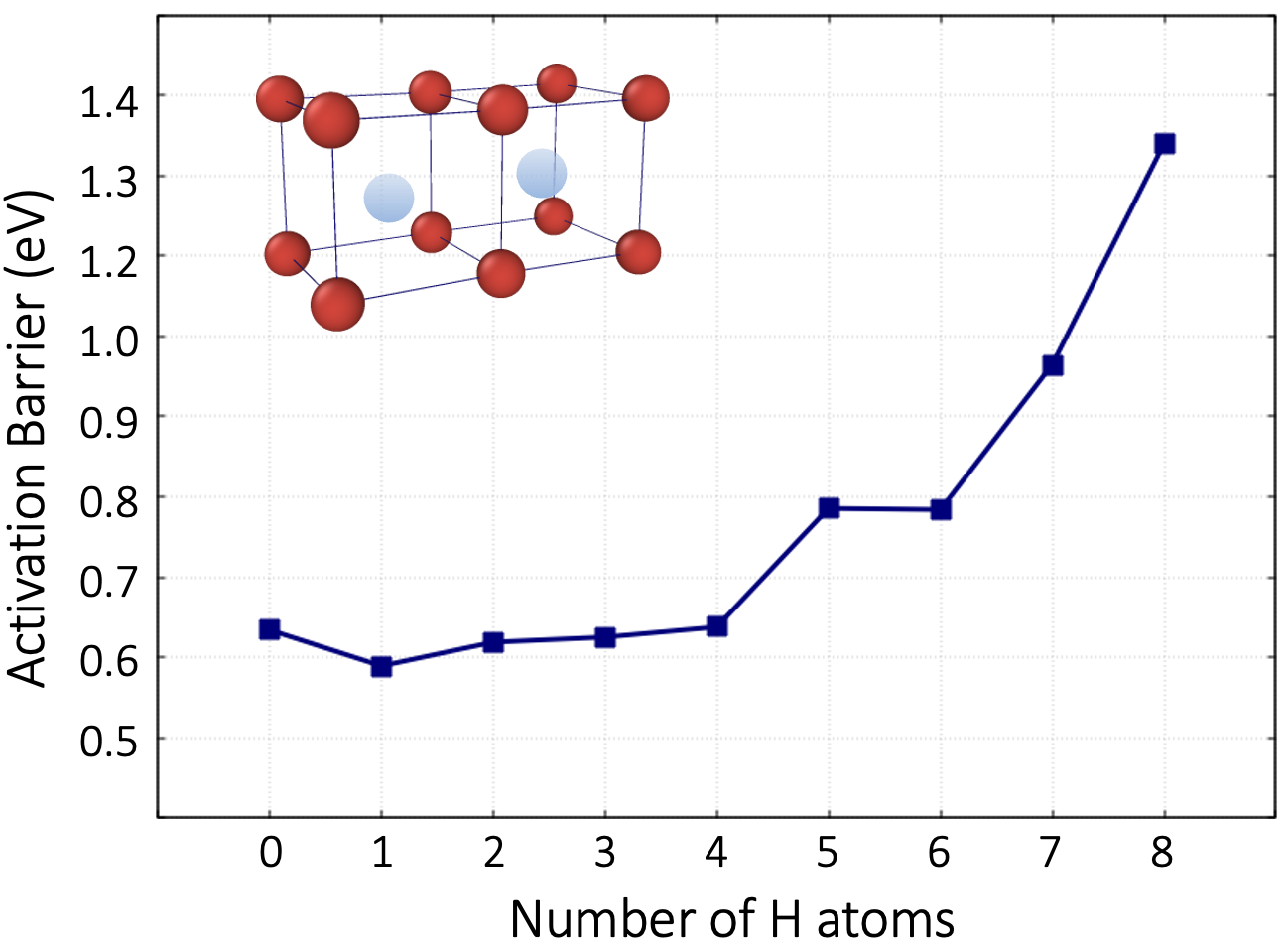}
    \caption{Smallest barrier for the first diffusion step into a EAM-characteristic split vacancy for any of the 12 Fe atoms next to the divacancy as a function of the number of trapped H atoms (V$_2$H$_x$).Inset: configuration of the divacancy with $x=0$. Gray spheres: vacancy sites; red spheres: Fe.}
    \label{fig11}
\end{figure}

To understand how H and vacancies interact as more vacancies aggregate, we turn to the case of two vacancies placed in the second neighbor position along the 100 direction as shown in Fig.~\ref{fig11}(inset). This orientation has the lowest formation energy than for divacancies in the first-neighbor position (111 direction). Launching kART from this position, we find that the two vacancies move together in a 4-step motion, crossing an overall 0.63~eV barrier. 

From this structure, we then look at the barriers associated with the first step of Fe diffusion into one of the vacancies in the presence of zero to 8 trapped H (V$_2$H$_x$) (Fig.~\ref{fig11}). To simplify the analysis, we started in all cases from the H configuration with the lowest energy. For $x=1$ to 4, the first step diffusion barrier is almost independent of the number of trapped H, going from 0.589 to 0.637~eV. Starting with $x=5$, the barrier goes up from 0.785~eV for $x=5$ and 1.23~eV for $x=8$.  As for the monovacancy, therefore, H does contribute to pinning the divacancy.

\section{\label{discussion}Discussion and Conclusion}
\label{section:Discussion}

In this work we use the kinetic Activation-Relaxation technique  (kART) to study vacancy plus H complexes (VH$_x$ and V$_2$H$_x$) in bcc Fe using a specifically developed EAM empirical potential~\cite{song2013atomic,ramasubramaniam2010erratum}.

As expected, k-ART reproduces known literature results regarding the Fe-H potential for the diffusion of a single H atom in a perfect BCC iron, as the H to jump from one four T-site to each of the four nearest T-site, with a prefactor, obtained using the harmonic transition theory, of $2.22 \times 10^{13}\, s^{-1}$ and a 0.04~eV barrier. Interestingly, while we could expect a lighter atom, such as H, to oscillate quickly, the prefactor is near standard values for the diffusion of heavier atoms~\cite{restrepo2016diffusion,restrepo2018carbon}.
Our simulations identify all the potentially favorable jumps for the H atom as well as the geometrical details that characterize the associated diffusion pathway.

Having established the consistency of EAM results with DFT and experiments, we turn to the interactions between H and vacancies. We start by characterizing the trapping of H by a vacancy. We find a strong binding energy for one H in a vacancy, at 0.603~eV, in agreement with DFT (0.559~eV and 0.60)~\cite{tateyama2003stability, mirzaev2014hydrogen}  and previous EAM calculations (0.603~eV) ~\cite{restrepo2020effect}. As the vacancy cavity is progressively filled with H atoms, the configuration becomes increasingly unstable until H atoms are ejected from it without crossing an energy barrier. This point of instability is reached when we try to insert 6 or more H atoms. This reproduces previous EAM calculations~\cite{restrepo2020effect}, as more H atoms (1 to 4) lead to a decrease in incremental binding energy, as shown in Table~\ref{tab:table1}. However, DFT predicts for the 5th H the binding energy will increase again (0.335~eV) contrary to our results (0.079~eV). This discrepancy could be due to small cells for DFT, as elastic deformation of the lattice becomes more important with a large number of trapped H atoms, or flaws in the EAM description of an H-dense environment.

We then focus on a detailed classification of the VH$_x$ complex diffusion pathways and assess how H changes the vacancy's diffusion mechanisms, starting with VH$_1$ and working our way up to five trapped H. As discussed in the results section, to find the diffusion pathways, we do both K-ART and ab initio-ARTn simulations for the VH$_x$ complex. We generate a  comprehensive description of the diffusion pathways and energy barriers. For VH$_1$, kART finds seven different activated events (this includes events where the H moves inside or leaves the vacancy). Based on this extensive catalog, two relevant full diffusion pathways are reconstructed, including the lowest-energy path leading to VH$_1$ complex diffusion. This lowest energy path follows an asymmetric diffusion trajectory where, first, a Fe moves over a 0.679~eV barrier to form a EAM-characteristic metastable split vacancy, 0.58~eV above ground state (Fig.~\ref{fig6}). The vacancy is reunited once the system moves a 0.172~eV barrier, 0.532~eV above ground state. In this configuration, the H finds itself in a metastable quasi-octahedral interstitial site outside of the vacancy and requires three further steps to move into the displaced vacancy site with a barrier of 0.02~eV (0.54~eV inverse barrier). Overall, therefore, the total effective diffusion energy barrier for the VH$_1$ complex is 0.758~eV. Except for the known metastable state EAM-characteristic artifact, this pathway is similar to that generated using ab initio-ARTn. Calculations show that it takes place through a vacancy jump with a barrier of 0.818~eV followed by a jump of the H atom to trap again inside the vacancy (Fig.~\ref{fig7}) consistent with an ab initio study by  Hayward~\cite{hayward2013interplay} who find that a 0.76~eV barrier when corrected for zero-point energy (0.79~eV without correction for zero-point energy).

This is the same mechanism as the one found by Restrepo \textit{et al.} (with the 0.759~eV barrier)~\cite{restrepo2020effect} in a simulation of vacancy diffusion in the presence of H in BCC Fe using parallel replica dynamics (PRD), the nudged elastic band method (NEB), and an analytical model by calculating minimum energy paths of migration. However, Restrepo and colleagues suggest rather that the reverse path is most favorable, with H moving out first. This reverse mechanism (see \Cref{fig6}) has, of course, the same overall diffusion barrier as the forward mechanism. However, even though the barrier to removing a H (first step) is lower, it is less likely, from a kinetic point of view: the need to cross multiple metastable steps before reaching the maximum barrier should make it less likely to occur than with the vacancy moving first. Clearly, further analysis is needed to quantify this question.

We note that a more straightforward jump, with H detraping first, is also observed in our simulation. However, it requires crossing a 1.104~eV effective barrier, corresponding essentially to the H detraping barrier plus that of a vacancy diffusion, making this mechanism much less probable than with the vacancy moving first. 

For the VH$_2$ complex, diffusion can take place mainly through two mechanisms:  with the vacancy dragging the H atoms (VH$_2$) together for diffusion, or with the unbinding one of the H atoms first (VH$_1$,1H), letting the VH$_1$ move to an adjacent site and reinserting H inside it (VH$_2$). For this complex, the diffusion of the vacancy ahead of the H (first mechanism) is the most probable, closely followed by a mechanism led by H detrapping. The first mechanism, shown in Fig.~\ref{fig8}, involves the production of a split vacancy, after crossing a 0.75~eV  barrier, followed by the completing of the vacancy jump through a 0.236~eV barrier. This brings the system in a metastable state (0.899~eV above ground state), with the vacancy having jumped one lattice site, leaving behind the two H. In three steps, one of the H moves into the vacancy (0.448~eV above ground state), followed, in four steps, by the second H. The total effective of this system is equal to 0.978~eV, higher than for the VH$_1$ complex. As discussed above, starting with first unbinding a H raises the overall barrier to 1.09~eV, presenting a similar probability.

This conclusion is different from the one presented by research~\cite{restrepo2020effect} for the diffusion of the VH$_2$ complex, based on the identification of a 0.743~eV barrier. In the same spirit as for the VH$_1$ complex, Retrespo and colleagues suggest that an H atom detraps from the vacancy first, followed by a Fe atom jumping into it.  However, this underestimates the overall barrier as it misses the second step, which involves crossing a 0.62~eV barrier, and, instead suggests a direct move from state 2 to state 4 (see \Cref{fig9}). As shown in Fig.~\ref{fig9} the H atom first detraps from the VH$_2$ with a barrier of 0.502~eV (VH$_1$, 1H), then the Fe atom jumps to the split vacancy site at 0.62~eV, resulting in a metastable position. A second energy barrier in Fe diffusion is 0.169~eV, then the Fe jumping from metastable to the ground position (1.099~eV total energy barrier). H atoms behind the vacancy are diffused to come around and trap inside the vacancy, resulting in steps 8-14 in Fig~\ref{fig9}.
Although this difference could be due to the EAM potential that we use in our simulation, the missing split interstitial in Ref.~\cite{restrepo2020effect} suggests that the pathway was not fully reconstructed. Our results are also compatible with a general trend obtained by looking at all mechanisms for the diffusion of VH$_x$ complexes, with $x=1$ to 5. As expected from the binding energy, the V+$x$H complex diffusion barrier increases with $x$, indicating that adding more H leads to stronger trapping of vacancies. 

As shown by Hayward and Fu ~\cite{hayward2013interplay},  zero point vibrational energy (ZPE) corrections reduce the trapping energies by about 0.11~eV for VH$_1$ and VH$_2$ complexes and by 0.04~eV for VH$_3$ and VH$_4$ complexes and almost zero for VH$_5$ complex. Although ZPE corrections may play an important role in the motions of H atoms inside vacancies, their effect is small for trapping and diffusion pathways of VH complexes.
The diffusion of the divacancy is also examined in the presence of H atoms (V$_2$H$_x$ complex) as well as in the absence of H atoms in this work. The barrier for the first step in the diffusion of a divacancy is 0.63~eV (consistent with KMC studies 0.62~eV~\cite{fu2005multiscale}), while the first barrier increases up to 1.23~eV once we have eight H atoms. This pinning is weaker than for a monovacancy with H: as shown in~\Cref{fig10,fig11} in the case of 1 vacancy diffusion, the barrier equals 0.64~eV, and adding first H will raise the barrier to 0.679~eV (for the first barrier of diffusion mechanism) and 1.602~eV for the case of 5H. That suggests that the pinning efficiency of H is reduced as the size of the vacancy cluster increases. Since these large clusters diffuse less, this effect is, nevertheless, decreasing significantly with the vacancy cluster size.

Overall, therefore, H atoms trapped in vacancies tend to slow their diffusion, in agreement with recent experimental study~\cite{chiari2021strain} that has shown that hydrogen trapping stabilizes vacancies and inhibits their diffusion, resulting in hydrogen embrittlement. An iterative process can result where the trapped hydrogen is released by vacancies, resulting in the diffusion of the vacancies, before being trapped again, as H diffuses quickly through the lattice. Transmission X-ray microscopy (TXM) results~\cite{lee2023detection} suggest that the nanovoids are not able to grow  to sizes larger than 100 nm when voids are stabilized by H, supporting our observation. As demonstrated as well in ~\cite{neeraj2017hydrogen}, in a variety of test conditions, high-resolution SEM studies of quasi-brittle fracture surfaces showed that fracture occurs as a result of nanovoids in the presence of hydrogen.
Numerical studies also show that that the hydrogen-vacancy complex is thermally stable and has low diffusivity~\cite{li2015interaction,tehranchi2016hydrogen}. Referencing the research~\cite{li2015interaction}, investigated the role of hydrogen-vacancy complexes in nucleating and growing proto nanovoids upon dislocation plasticity in BCC Fe by molecular dynamics and cluster dynamics simulations. V$n$H$_x$ complexes are thermally stable, according to them.
Earlier studies have consistently found that hydrogen-induced vacancies in materials are strongly linked to premature fractures caused by hydrogen exposure~\cite{sugita2021review}. These vacancy-based mechanisms contribute to the development of cracks in Fe, thereby embrittling it. A significant amount of H can be trapped at vacancies, leading to an increase in local H concentration without the likelihood of H accumulating in bulk interstitial sites. It is possible to achieve a sufficiently high H concentration in Fe, which has a very low equilibrium H concentration in bulk, to successfully induce H embrittlement.

It remains unclear why and how nanovoids, detected behind fracture surfaces, contribute to HE, as observed in other experimental studies~\cite{neeraj2012hydrogen,neeraj2017hydrogen}. Further research is needed in order to answer this question with regard to vacancy clustering, as well as the interaction of V$n$H$_x$ complexes with other defects such as grain boundaries and dislocations.

As a result, k-ART provides a comprehensive and detailed understanding of all diffusion mechanisms allowing both mechanistic insights into the embrittlement process and quantitative prediction capabilities. We are also able to support the k-ART results with DFT calculations. Beyond what was discussed, k-ART has great potential for the future of modeling H-defect interactions and improving our understanding of HE on an atomic scale, which will lead to the development of H-embrittlement resistant materials.

\section{Code availability}

The k-ART and ab initio-ARTn packages are freely available upon request. Please contact the authors to have access to the repository. 

\section{\label{sec:level1}Acknowledgments}

This work is supported in part by a grant from the Natural Sciences and Engineering Research Council of Canada (NSERC). We are grateful to Calcul Québec and Compute Canada for providing generous computer resources. This study was facilitated by the powerful OVITO software developed by A Stukowski ~\cite{stukowski2009visualization}, which assisted in analyzing and illustrating atomic configurations. You can access OVITO at http://ovito.org/. 

\bibliography{apssamp}
\bibliographystyle{apsrev4-1}

\end{document}